\documentclass[aps, prl, twocolumn, superscriptaddress, amsmath,  tightenlines, longbibliography]{revtex4-2}
\usepackage{graphicx}
\usepackage{dcolumn}
\usepackage{bm}
\usepackage{bbm}
\usepackage{framed} 
\usepackage{tcolorbox}
\usepackage{braket}
\usepackage{float}	
\usepackage{chapterbib}
\usepackage{amsthm,amsmath,amssymb}
\usepackage{mathrsfs}
\usepackage{amsfonts}	
\usepackage{epsfig}
\usepackage{multirow}	
\usepackage{booktabs}
\usepackage{color}
\usepackage{xcolor}

\newcommand{\nvec}[1]{\textbf{#1}}

\newcommand{\im}{i}
\newcommand{\ii}{\im}

\newcommand{\e}{e}

\def\ie{{\it i.e.\/}}

\begin{document}
\preprint{APS/123-QED}

\title{Evidencing non-Bloch dynamics in temporal topolectrical circuits}
\author{Maopeng Wu}
	\affiliation{State Key Laboratory of Tribology, Department of Mechanical Engineering,\\
Tsinghua University, Beijing 100084, China}	
\affiliation{Department of Engineering Physics, Tsinghua University, Beijing 100084, China}	
\author{Qian Zhao}%
 	\email{zhaoqian@tsinghua.edu.cn}
 	\affiliation{State Key Laboratory of Tribology, Department of Mechanical Engineering,\\
Tsinghua University, Beijing 100084, China}
\author{Lei Kang}
	\affiliation{Department of Electrical Engineering, \\
The Pennsylvania State University, University Park, PA 16802, USA}	
\author{Mingze Weng}
	\affiliation{State Key Laboratory of Tribology, Department of Mechanical Engineering,\\
Tsinghua University, Beijing 100084, China}	
\author{Zhonghai Chi}
	\affiliation{State Key Laboratory of Tribology, Department of Mechanical Engineering,\\
Tsinghua University, Beijing 100084, China}	
\author{Ruiguang Peng}
	\affiliation{State Key Laboratory of Tribology, Department of Mechanical Engineering,\\
Tsinghua University, Beijing 100084, China}	
\author{Jingquan Liu}
	\affiliation{Department of Engineering Physics, Tsinghua University, Beijing 100084, China}	

\author{Douglas H. Werner}
	\affiliation{Department of Electrical Engineering, \\
The Pennsylvania State University, University Park, PA 16802, USA}	

\author{Yonggang Meng}
	\affiliation{State Key Laboratory of Tribology, Department of Mechanical Engineering,\\
Tsinghua University, Beijing 100084, China}	
\author{Ji Zhou}
	\affiliation{State Key Laboratory of New Ceramics and Fine Processing,\\
School of Materials Science and Engineering, Tsinghua University, Beijing 100084, China}	
\date{\today}

\begin{abstract}

One of the core concepts from the non-Hermitian skin effect is the extended complex wavevectors (CW) in the generalized Brillouin zone (GBZ), while the origin of CW remains elusive, and further experimental demonstration of GBZ is still lacking. We show that the bulk states of an open quantum system dynamically governed by the Lindblad master equation exhibit non-Bloch evolution which results in CW.
Experimentally, we present temporal topolectrical circuits to serve as simulators for the dynamics of an open system. By reconstructing the correspondence between the bulk states of an open system and circuit voltage modes through gauge scale potentials in the circuit, the non-Bloch evolution is demonstrated. Facilitated by the simulators and proper approach to characterize the non-Bloch band proposed here, the GBZ is confirmed. Our work may advance the investigation of the dissipative topological modes and provide a versatile platform for exploring the unique evolution and topology for both closed and open systems.
\end{abstract}

\maketitle

\textit{Introduction}.---In topological matter \cite{hasan2010colloquium,qi2011topological,lu2014topological,ozawa2019topological}, the number of topological surface states is usually related to the topological structure of bulk states (or topological invariances), which is summarized
as the bulk-boundary correspondences (BBC). Recently, those topological phase studies have been generalized to the systems with non-Hermitian (NH) Hamiltonians \cite{bergholtz2021exceptional}, which usually root in the intrinsic dissipative property (e.g., inelastic-scattering-induced finite quasiparticle lifetime \cite{kozii2017non}, material gain/loss or resonantor Q factor \cite{feng2017non}) or the interaction with the surrounding environments (e.g., Liouvillian dissipators \cite{diehl2011topology}). The interplay between non Hermiticity and topological phases has attracted intense interest not only because of the exotic phenomena, such as the gain/loss induced band attraction \cite{wu2020energy}, but also because of the potential for exciting applications, such as the NH laser \cite{qiao2021higher}.

Nevertheless, recent studies show that reported BBC theories for Hermitian systems are not well defined in NH systems. Subsequent investigations revealed that a broad range of NH models inherently exhibits the non-Hermitian skin effect (NHSE) \cite{xiong2018does,yao2018edge,longhi2019probing,zhang2020correspondence,okuma2020topological}, which can be captured mathematically by complex wavevectors (CW). Namely, the bulk eigenstates of the NH Hamiltonian do not extend over the systems but prefer exponentially localizing (or piling up) at its boundaries. Accordingly, the Bloch theorem is generalized to $\psi_{\tilde{\mathbf{k}} } \left (\mathbf{r}  \right ) =e^{i\tilde{\mathbf{k}}\cdot \mathbf{r}}u_{\tilde{\mathbf{k}} }\left ( \mathbf{r}\right )$, where $\tilde{\mathbf{k}} $ belongs to the generalized Brillouin zone (GBZ). The GBZ provides an amendment deals with anomalous BBC by replacing $e^{i\mathbf{k}}$ with $e^{i\tilde{\mathbf{k}}}$  in calculating the topological invariants, e.g., paradigmatic one-dimensional NH SSH model \cite{yao2018edge,jin2019bulk} as experimentlly demonstrated in \cite{xiao2020non}, NH Chern insulators \cite{yao2018non}, NH second-order topological insulator \cite{liu2019second} and etc. However, the underlying physical mechanism of this phenomenon-driven development of CW remains elusive. Furthermore, owing to the lack of the appropriate approach to characterizing the non-Bloch band, the GBZ has not been experimentally demonstrated.

In this letter, we investigate a second-order topological insulator that interacts with the environment, thereby its time evolution follows the Lindblad master equation. We show the time evolution of the bulk field coherences experiences an additional space-related negative damping term beyond Bloch theorem (dubbed as the non-Bloch evolution), which leads to CW. Also, the concept of topolectrical circuits is adapted to emulate the non-Bloch evolution. By contrast, the gauge scale potentials in the proposed circuits bridge the bulk field coherences and the voltage modes, which lay the necessary basis for emulating. Moreover, we reveal that Laplace transform rather than Fourier transform connects the reciprocal space and the real lattice space for non-Bloch problems, indicating an unexplored avenue to characterize the non-Bloch band. The above revelation, together with the propoesd circuits, finally allows us to verify the existence of the GBZ experimentally. In separate work \cite{wusemi}, we apply temporal topolectrical circuits to directly evidence non-Abelian nodal-line semimetals \cite{wu2019non} as well as their unique earring nodal structure, which can not be reached by other platforms.
\par
\begin {figure*}
\centering
\includegraphics[width=13cm]{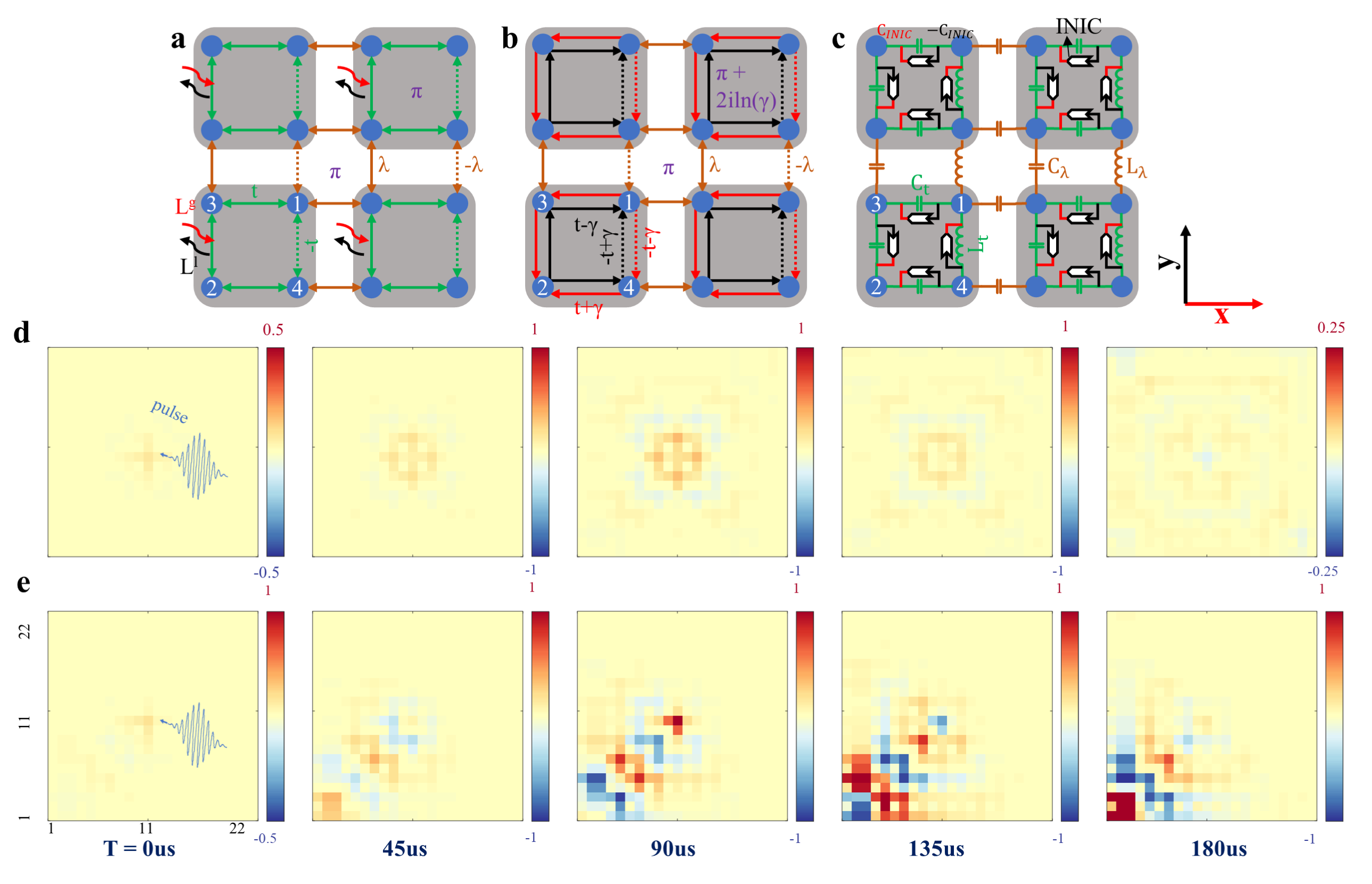}
\caption{Non-Bloch evolution and its demonstration using a temporal topolectrical circuit. (a)
Tight-binding representation of the model.
Each unit cell contains four orbitals (solid blue circles). The green and brown arrows denote the intercell and intracell hopping, respectively.
The dashed arrows have a relative negative sign to account for a flux of $\pi$ threading each plaquette. The dissipators $L^l$ and $L^g$ describe the boson loss and gain in a unit cell. (b) The effective Hamiltonian $H$ for $\phi_{\mathbf{k},i}\left ( t\right )$ in real space. The red and black arrows represent the asymmetric intracell hopping, accounting for an imaginary flux  $2iln(\gamma)$. (b) Temporal topolectrical circuit realization of the non-Bloch evolution. $(C_{t}, L_{t})=(10\  nF,30\  uH)$, $\lambda=2$, $\gamma=0.5$. The configuration of $Y_{0} = iw_0C_{0},\ C_{0}=33\ nF$ is not plotted for clarity. (c) Measured time-resolved voltage signal of the Hermitian topolectrical circuit ($11\times 11$ units) after the pulse. The node voltage of the bulk modes propagates in all directions and behaves as a Bloch wave with forming a quasi-cylindrical wavefront. (d) Time-resolved voltage signal of the NH circuit after pulse. In contrast, the node voltage propagates chirally toward the corner with the increasing amplitude, confirming the non-Bloch dynamics. Note that the chirality of the non-Bloch wave is controlled by the dissipators.}
\label{fig1}
\end{figure*}

\textit{Non-Bloch evolution in quantum open systems}.---Following the Lindblad master equation \cite{lindblad1976generators}, the time evolution of a reduced density matrix $\rho$ of an open system is governed by: 
\begin{equation}
\label{masterEq}
\frac{\mathrm{d}\rho }{\mathrm{d}t} = -i\left [ H_0, \rho \right ]+ \sum _{u}\left ( 2L_u \rho L_u^{\dagger} -\left \{L_u^{\dagger}L_u,\rho \right \}\right ),
\end{equation}
where $H_0$ is the Hermitian Hamiltonian of the original system. $L_u$ is the set of dissipators due to environment, and with the translational symmetry $L_u$ can be simplified as $L_u = \left \{L_{\mathbf{k}}^g, L_{\mathbf{k}}^l\right \}$ in BZ, which include the particle gain $L_{\mathbf{k}}^g=\psi_{\mathbf{k}}^{\dagger}D_{\mathbf{k}}^g$ and the loss $L_{\mathbf{k}}^l=D_{\mathbf{k}}^l\psi_{\mathbf{k}}$, $\psi_{\mathbf{k}} =\left ( a_{\mathbf{k},1},a_{\mathbf{k},2},...,a_{\mathbf{k},n}\right )^T$, $a_{\mathbf{k}, i}$ is the annihilation operator, $n$ represents the degrees-of-freedom per unit cell. We further assume the element constitutions are bosons which are distinct from the fermions in commutation relation \cite{song2019non}. The field coherences $\phi_{\mathbf{k},i}\left ( t\right )=\left \langle a_{\mathbf{k}, i}\left ( t\right )\right \rangle = Tr\left [ a_{\mathbf{k},i}\rho\left ( t\right )\right ]$ are employed to monitor the time evolution of $\rho$ because of its accessibility.
Equation \ref{masterEq} implies that 
$\phi_{\mathbf{k},i}\left ( t\right )$ evolve under an effective NH Hamiltonian $H$ \cite{wu}:
\begin{align}
&\frac{\mathrm{d}\phi_{\mathbf{k},i} }{\mathrm{d}t} = -i\sum_{m}H_{m,n}\phi_{\mathbf{k},j} \nonumber \\ 
&H = H_0+\frac{i}{2}\left ( \left ( D_{\mathbf{k}}^{g\dagger}D_{\mathbf{k}}^g\right )^T-D_{\mathbf{k}}^{l\dagger}D_{\mathbf{k}}^l\right ).
\end{align}
Above all, we have the eigenmodes of the field coherences $\phi_{\mathbf{k},n}\left ( t\right )$: 
\begin{equation}
\phi_{\mathbf{k},n}\left ( t\right )= \Phi_{\mathbf{k},n}e^{iE_n(\mathbf{k})t-i\mathbf{k}\cdot \mathbf{r}},
\label{psi}
\end{equation}
where $\Phi_{\mathbf{k},n}$ and $E_n(\mathbf{k})$ are the eigenvectors and eigenvalues of $H$, respectively. Notably, $H$ reduces to $H_0$ in the absence of the coupling.\par
Without loss of generality, we consider a 2D system consisting of four orbitals arranged on a square lattice (Fig. \ref{fig1}b). This non-dissipating model has been used to study the higher-order topological corner states \cite{benalcazar2017quantized,serra2018observation,peterson2018quantized}.
Heuristically, we note that, to construct an intriguing model that exhibits the non-Bloch evolution and hosts the topological corner states, the dissipators should have the following properties: (i) ensure $H$ is NH, $H \ne  H^\dagger $, (ii) introduce certain (\ie, asymmetric) hopping between the orbitals to enable non-Bloch behavior; and (iii) ensure $H$ respects certain symmetry to allow the topological phase. Regarding those properties, the dissipators considered here are 
\begin{equation}
D_{\mathbf{k}}^g = 2\sqrt{\gamma } \begin{pmatrix}
1 &  &  & \\ 
 &  1&  & \\ 
 &  &  1& \\ 
 &  &  & 1
\end{pmatrix},
D_{\mathbf{k}}^l = \sqrt{2\gamma } \begin{pmatrix}
1 &  &-i  & \\ 
 &  1&  &i \\ 
 & -i &  1& \\ 
 -i&  &  & 1
\end{pmatrix}.
\end{equation}
and therefore
\begin{equation}	
\begin{aligned}
\label{effH}
H = &\left[t + \lambda \cos(k_x)\right] \tau_x\sigma_0 - \left[\lambda \sin(k_x) ~ + i \gamma \right] \tau_y \sigma_z + \\ & \left[t + \lambda \cos(k_y)\right] \tau_y \sigma_y + \left[\lambda \sin(k_y) + i \gamma\right] \tau_y \sigma_x.
\end{aligned}
\end{equation}
where $(t,-t)$ and $(\lambda, -\lambda)$ are the intracell and intercell hopping, respectively. $\tau_i$ and $\sigma_i$ ($i = x,y,z$) are Pauli matrices. 
To present a physical picture of the effective NH Hamiltonian $H$, we transform it back into real space shown in Fig. \ref{fig1}b. The interactions $L_u$ between the original system and the environment can be interpreted as asymmetric intracell hopping, which accounts for a complex flux threading the unit cell.
The topological property in our model will be reported elsewhere.\par
Our calculation shows that \textit{bulk} field coherences of an excitation (like a photon) centered at the lattice prefer to evolve toward the specific direction, while evolution along the opposite direction is suppressed. In other words, field coherences experiences a negative space-related damping term during the evolution, which behaves as the increasing number of excitation along the direction it prefers. 
The damping term is independent of boundary condition, quite different from the eigenvalue spectrum that shows great sensitivity to the boundaries. Even so, the system with periodic boundary conditions never reaches a steady state as the field coherences will exponentially and constantly accumulate along the periodic direction. We remark that the CW caused by non-Bloch evolution is independent of quantum coherence, which gives rise to a possibility of emulating the evolution with classical circuits. In fact, the NHSE (described by CW) may dramatically shape the long-time Lindblad dynamics after the jump \cite{song2019non}.

\begin {figure}[t]
\centering
\includegraphics[width=\linewidth]{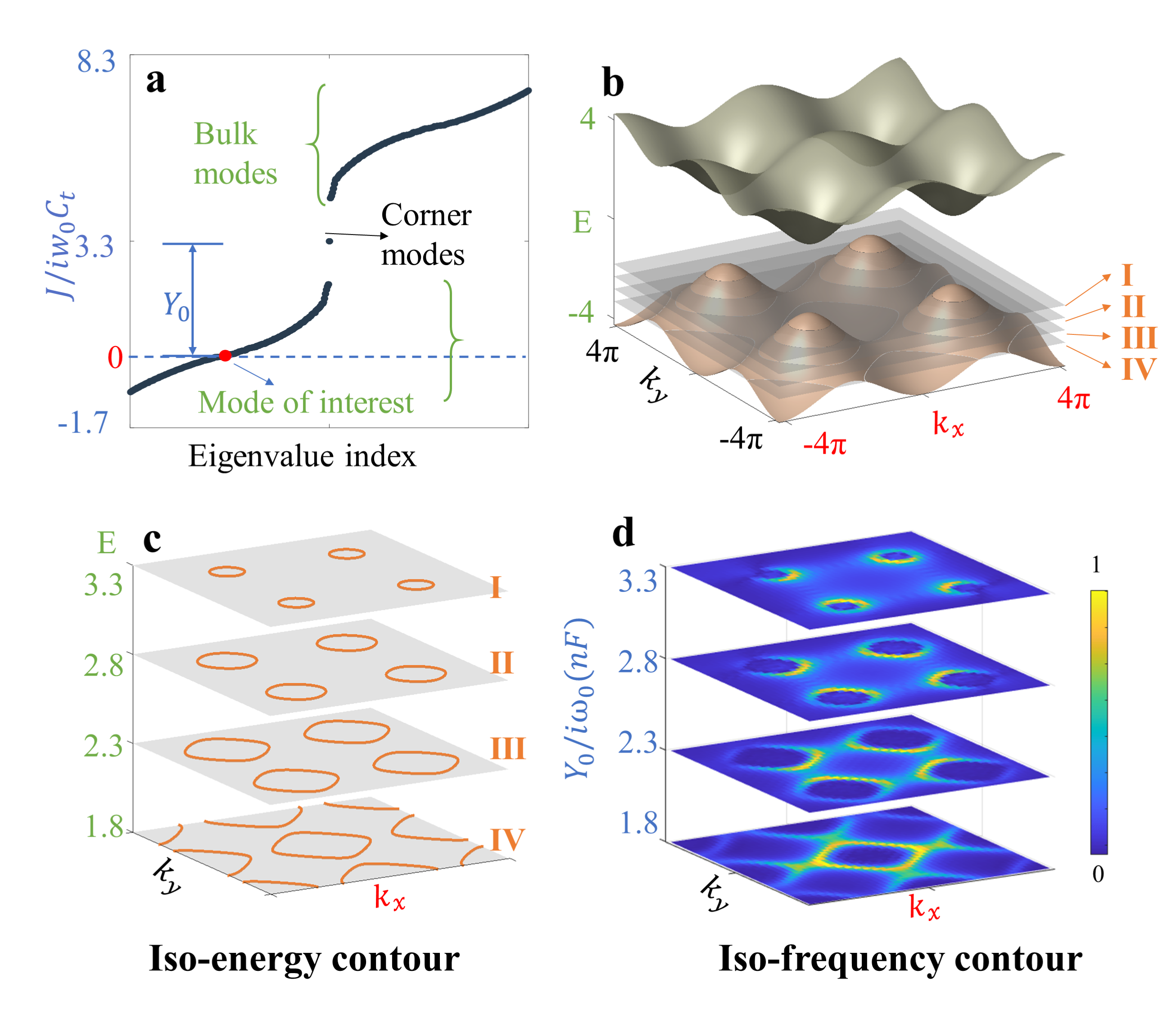}
\caption{Selecting time evolution pattern by adjusting the gauge scalar potentials $Y_0$ and it enabled FTFS for \textit{Bloch} band. (a) Eigenvalue spectrum of the circuit admittance matrix $J\left ( w, \mathbf{k}\right ))$. The red dot denotes the bulk modes whose eigenvalues are zero. The spectrum is gauge invariant against the diagonal term $Y_{0}\tau_0\sigma_0$ of $J\left ( w, \mathbf{k}\right )$, hence it shifts the spectrum upwards by $Y_0 = iw_0C_0, C_0 = 33nF$. (b)-(d) Charactering circuit band by tuning $Y_0$. (b) Band structure of $H$ with $\gamma = 0$ (Hermitian case), which corresponds to $J\left ( w, \mathbf{k}\right ))$. I-IV denote the cut plane with $E = 1.8, 2.3, 2.8$ and $3.3$ respectively. Their intersections with the band are the iso-energy contours, which are individually shown in (c). (d) shows voltage distribution of the Hermitian circuit (21 by 21 units) in BZ after the Fourier transform with $C_0 = 18nF, 23nF, 28nF$ and $33nF$  respectively (more details see \cite{wu}).}
\label{fig2}
\end {figure}
\textit{Emulating non-Bloch dynamics with temporal topolectrical circuits.}---For the circuit in a lattice structure \cite{cserti2011uniform}, its admittance matrix in BZ can be described as $J\left ( w, \mathbf{k}\right )  = YI + J_{0}\left ( w, \mathbf{k}\right ) $, where the diagonal elements $YI$ (so-called self admittances) are preset to be indentical. The node voltage $V_{\nvec{k},0}(t)$ that changes over time are \cite{wu}
\begin{align}
    V_{\nvec{k}}(t) = \nvec{V}_{\nvec{k},0} \, \e^{\ii \omega(\nvec{k}) t- \ii \nvec{k} \cdot \nvec{r}}, 
\label{volOverTime}
\end{align} 
where $\nvec{V}_{\nvec{k},0}$ are the eigenvectors of $J\left ( w, \mathbf{k}\right )$ with respect to the \textit{zero-value eigenvalues}, $w$ is the resonant frequency of the circuit. The above equation reminds us of Eq. 3, and indicates that the response of the node voltage can potentially emulate the evolution of the field coherences, as long as $J\left ( w, \mathbf{k}\right )$ is appropriately configured to be of the same form as that of $H$ (a little different from Eq. 3, and we will solve this with gauge scale potentials). \par

\begin {figure}[t]
\centering
\includegraphics[width=7cm]{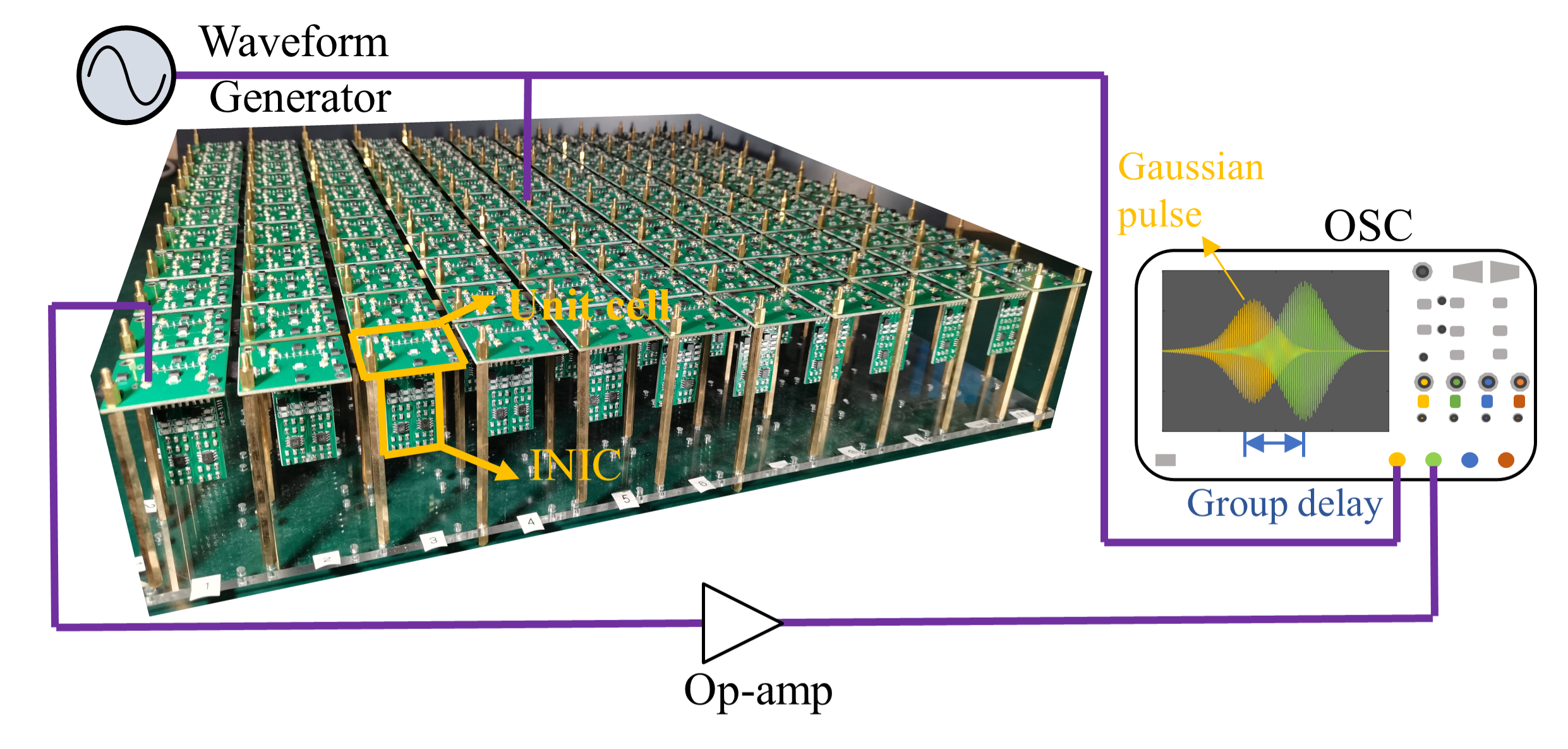}
\caption{ Schematic of the measurement setup with a photo of
the fabricated circuit. 
Orange rectangles denote a single cell module emulating the original Hermitian system and an INIC module dominating the coupling between the systems and the environment, respectively. 
Each module is assembled by pin header connectors. Note that the modular
design strategy dramatically facilitates circuit debugging and
calibration. The oscilloscope (OSC) records the node voltage variation over time.}
\label{fig3}
\end {figure}

The unit cell of the circuit designed to emulate $\phi_{\mathbf{k},i}\left ( t\right )$ is specified in Fig. \ref{fig1}c. Capacitors-inductors pairs ($C_{t}$, $L_{t}$) and ($C_{\lambda}$, $L_{\lambda}$) are used to emulate the intracell and intercell hopping of $H$.
$C_{\lambda}  = \lambda C_{t}$ and $L_{\lambda}  = L_{t}/\lambda$ so that the pairs have the same $LC$ resonant frequency $w_0$.
Furthermore, we implement the negative impedance converters with current inversion (INIC), which served as an indictor of the coupling between the system and the environment \cite{wu}. 
Each node is grounded via passive elements to guarantee the self admittance $Y=Y_{0}\tau_0\sigma_0$. The admittance matrix $J\left ( w_0, \mathbf{k}\right )$ of the resulting circuit reads  
\begin{equation}
\begin{aligned}
\left ( iw_0 C_{t}\right )^{-1} & J\left ( w_0, \mathbf{k}\right )    = Y_{0}\tau_0\sigma_0+ H
\label{admitMatrix}
\end{aligned}
\end{equation}
Neglecting the constant prefactor, $H$ and $J\left ( w_0, \mathbf{k}\right )$ take the same form when $Y_{0}\tau_0\sigma_0=0$, nevertheless as we will see, $Y_{0}\tau_0\sigma_0 \ne 0$ is essential in our temporal circuit.\par

The existing topolectrical circuit has severe limitations preventing us from exploring non-Bloch evolution and further GBZ.
For normal metamaterials or photonic crystals \cite{lu2014topological,ozawa2019topological,zhao2009mie,peng2017temperature}, the frequency dimension plays a similar role as the eigenenergy of a solid. However, to keep
$J\left ( w_0, \mathbf{k}\right )$ the same form as $H$, the circuit has to operate at $w_0$. Besides, the voltage response $V_{\nvec{k}}(t)$ (emulating $\phi_{\mathbf{k},n}\left ( t\right )$) is modeled by $\nvec{V}_{\nvec{k},0}$ with vanishing eigenvalue (Eq. 6), while in most cases, the eigenvalue of $\phi_{\mathbf{k},n}\left ( t\right )$ usually is not zero. Here, we find $Y_{0}$ functions as the gauge scalar potentials \cite{wu}. So, one can tune the eigenvalue spectrum with the $Y_0$ (Fig. 2a), just like tuning Fermi energy with chemical potentials. In other words, one can select the evolution of the bulk field coherences $\phi_{\mathbf{k},n}\left ( t\right )$ with tuning $Y_0$. As we will see, $Y_0$ does more than selecting modes, and it also plays an essential role in GBZ demonstration.

\begin {figure}[h]
\centering
\includegraphics[width=7cm]{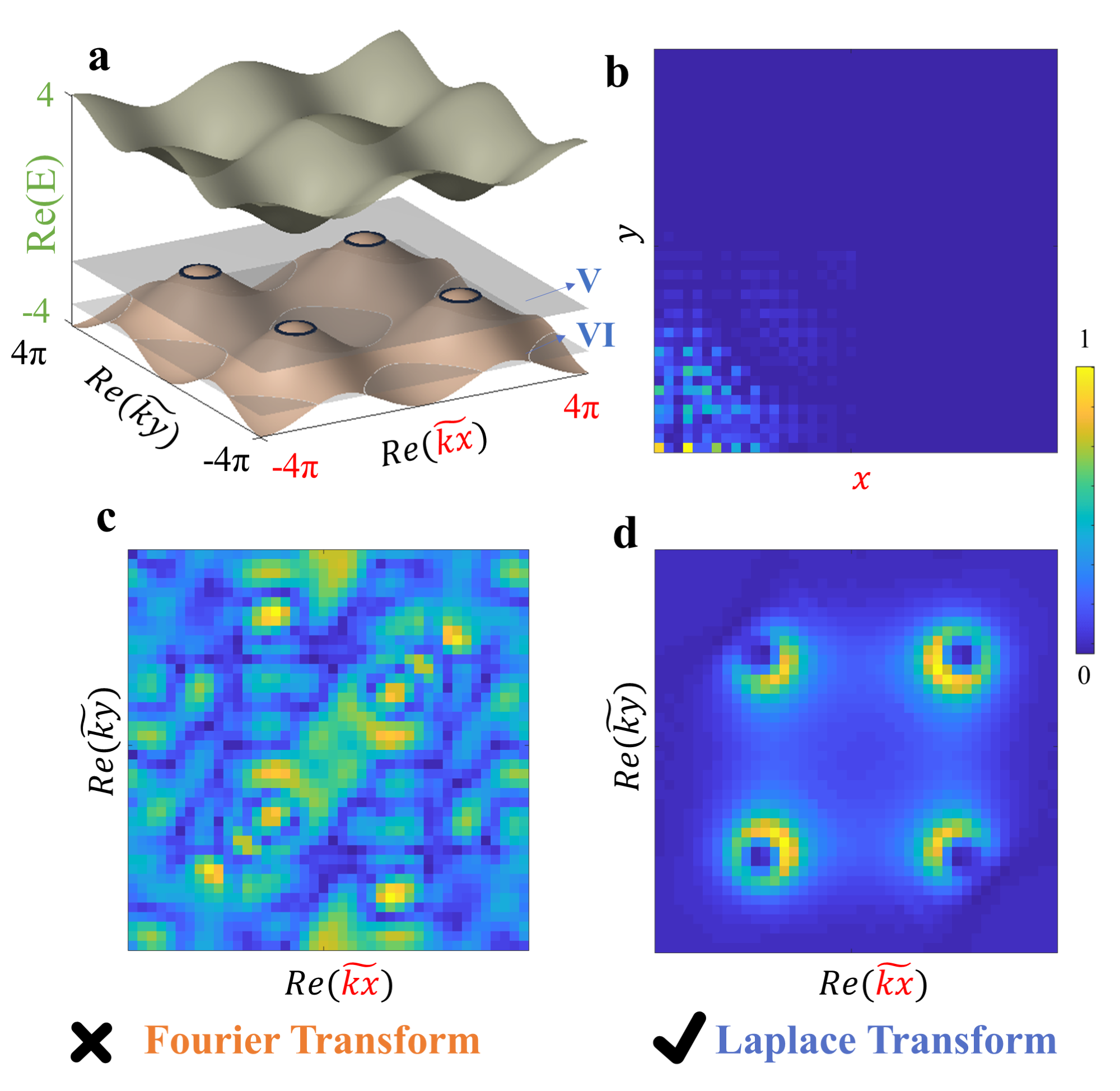}
\caption{Failture of the FTFS for non-Bloch band and characterizing the non-Bloch band with Laplace transform. (a) Non-Bloch band of H with $\gamma = 0.5$. V and VI denote the cut
plane with $E = 1.8$ and $3.3$ respectively. (b) Calculated voltage response due to a centered point source (not shown) in real space ($C_0 = 18nF$).
(c) Corresponding voltage distribution in BZ after the Fourier transform. Clearly, Fourier transform cannot be used to characterize the non-Bloch band.
(d) Corresponding voltage distribution in GBZ after the Laplace transform. The distribution is in great accord with the iso-contour (V in (a)). As for VI, see Fig. \ref{fig5}d}
\label{fig4}
\end {figure}

\begin {figure}[t]
\centering
\includegraphics[width=7cm]{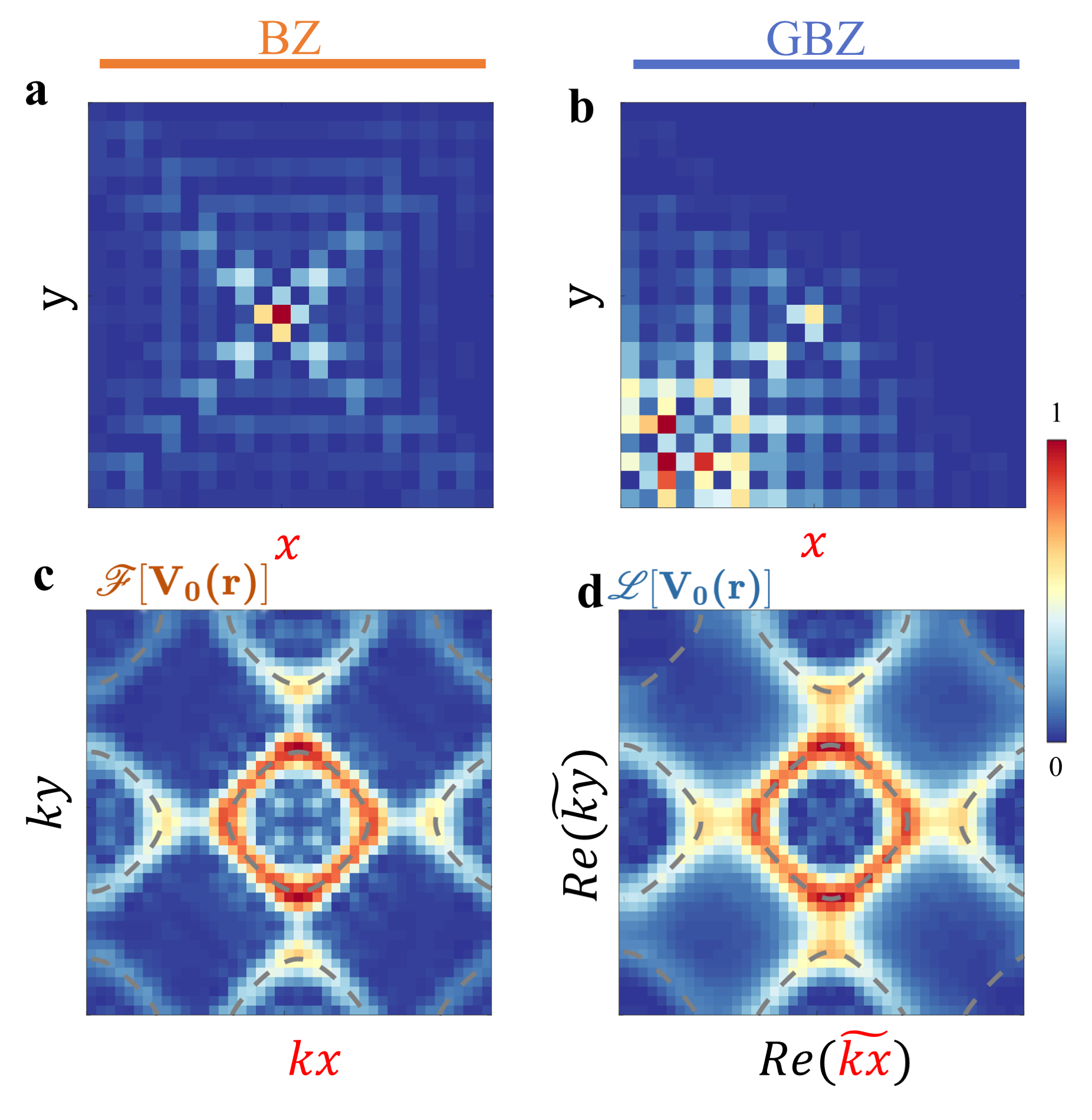}
\caption{Experimental demonstration of GBZ. (a) and (c) Measured voltage distribution (due to a centered point source)
of the Hermitian system in real space and its iso-frequency
contour after Fourier transform. (b) and (d) Measured voltage distribution of the NH system in real space and its iso-frequency contour after the Laplace transform. The voltage
distribution in (d) is localized at the corner as a result of the
NHSE. Gray dash lines indicate the calculated contour. }
\label{fig5}
\end {figure}

Experimentally, to excite an \textit{bulk} excitation, a current pulse signal with Gaussian form is injected into the center of the circuit (Fig. 3). For direct comparison, the same experiment
was also carried out in a Hermitian circuit. Clearly, in the NH system (Fig. 1e) the field coherences represented by the voltage response propagate toward the bottom-left corner of the lattice
with a negative damping factor (the amplitude increases as a function of the distance). This is in sharp contrast to the symmetric propagation phenomenon (referred as Bloch evolution) observed in a normal Hermitian system (Fig. 1d).\par
\textit{GBZ demonstration.}--- 
The Gauge scalar potentials $Y_0$ also enables an alternating approach to character the circuit band, opening up possibilities for the demonstration of GBZ. In traditional circuits, the circuit bands are obtained through the (or block) diagonalization of the measured impedance matrix. The diagonalization approach has tremendous convenience in studying the sensitivity of the spectrums \cite{imhof2018topolectrical,helbig2020generalized}, but it fails in demonstrating GBZ as GBZ is a prerequisite in the diagonalization process. The desired approach is the Fourier-transformed field scan (FTFS) if the field distribution (e.g., field coherences) is accessible. FTFS is widely used in artificial crystals and metamaterials, such as \cite{de2005fourier,yang2018ideal,yang2019realization}. As mentioned above, the bulk field distribution is accessible and can be selected at will by adjusting $Y_0$, thus enabling FTFS to character the \textit{Bloch} circuit band (Fig. 2b-2d).\par

FTFS is not applicable for the \textit{non-Bloch} band. Without NHSE, FTFS allows one to retrieve the band structure by applying the Fourier transform to the measured spatial field patterns (see Fig. 2b-2d, to avoid NHSE we take $\gamma = 0$).  Because the degeneracy modes with identical eigenenergy are simultaneously excited, the obtained distribution in momentum space is the iso-energy contour of the band. However, Fourier transforms of the non-Bloch wave are divergent when FTFS is applyed (Fig. 4c). This divergence arises from the limitation of the Dirichlet conditions \cite{debnath2016integral}. Therefore, FTFS is still far from demonstrating GBZ.

For non-Bloch problems, we find that the the Laplace transform rather than Fourier transform can map eigenvectors in real space to that in GBZ (Fig. 4d). In order to make a convergent transformation, an exponential decay (or growth) factor is usually introduced to the integral kennel, leading to the $e^{i\tilde{\mathbf{k} } \cdot \mathbf{r}}\rightarrow e^{\mathbf{s} \cdot \mathbf{r}}, \mathbf{s}\subset \mathbb{C} $. This is well-known as the bilateral Laplace transformation. In our generalized theory, we interpret the complex argument $\mathbf{s}$ is as CW, then accordingly, the Laplace transformation decomposes the wave
function $\psi_n\left (\mathbf{r} \right )$ in real space into that in GBZ:
\begin{equation}
\psi_n'\left ( \tilde{\mathbf{k} } \right ) = \mathcal{L}\left [ \psi_n\left (\mathbf{r} \right )\right ] =\int _{\mathbf{r} }\psi_n e^{i\tilde{\mathbf{k} } \cdot \mathbf{r}}d \mathbf{r}
\end{equation}
Fig. 5b and 5d present the measured voltage distribution and the voltage profile after the Laplace transformation, which is in great agreement with the theoretical analysis (more data about GBZ demonstration, see \cite{wu}). As a control for GBZ, we also measure the voltage field distribution of the Hermitian systems ($\gamma=0$) and transform it into momentum space using the Fourier transform (Fig. 5a and 5c). Clearly, GBZ is experimentally demonstrated. 
Here, the $H\left ( \tilde{\mathbf{k} } \right )$ in GBZ is assumed in the calculation of the non-Bloch band shown in Fig. 4a. This assumption is based on real space and guarantees the energy levels are continuous for a large open lattice \cite{yokomizo2019non}.

\textit{Conclusion.}---Non-Hermitian systems promise a pathway to expand the parameter space (such as complex
wavevectors) into the complex realm. Here, we show that the non-Bloch evolution of the open systems can result in complex wavevectors and experimentally demonstrate the GBZ, which accommodates the complex wavevectors. The temporal topolectrical circuit developed in our work provides a simple-to-realize platform to study the dynamics and topology of open quantum systems (and of course closed systems, such as \cite{wusemi}). Our works could advance the application of topological systems, such as topological lasers whose topological lasing modes extend over the bulk or directional amplification with excellent isolation.\par
\textit{Acknowledgments}---The authors thank Zhong Wang,
Fuli Zhang, Yuancheng Fan, Nianhai Shen, and Peng
Zhang for useful discussions and suggestions.
Funding:
This work is supported by the National Natu-
ral Science Foundation of China (51872154), the Bei-
jing Municipal Science and Technology Commission
(Z191100004819001).



\end{document}